\documentclass[aps,twocolumn,prl,showpacs,showkeys,nofootinbib,superscriptaddress]{revtex4-1}
\usepackage{amsmath, amssymb, amsfonts}
\usepackage[margin=1in]{geometry}
\usepackage{graphicx}
\usepackage{color}
\usepackage{xcolor}
\usepackage{lineno}
\usepackage{bm}
\usepackage{ulem}
\pagecolor{white}

\newcommand{\bea}{\begin{eqnarray}}
\newcommand{\eea}{\end{eqnarray}}
\newcommand{\beq}{\begin{equation}}
\newcommand{\eeq}{\end{equation}}
\newcommand{\bqa}{\begin{eqnarray}}
\newcommand{\eqa}{\end{eqnarray}}

\graphicspath{{Figures/}}

\begin{document}
\title{Explaining Snowball-in-hell Phenomena in Heavy-ion Collisions\\ 
Using a Novel Thermodynamic Variable}
\author{Eric Braaten}
\email{braaten.1@osu.edu}
\affiliation{Department of Physics,
	         The Ohio State University, Columbus, OH\ 43210, USA}
	
\author{Kevin Ingles}
\email{kingles@illinois.edu}
\affiliation{Department of Physics,
	         The Ohio State University, Columbus, OH\ 43210, USA}
\affiliation{%
Illinois Center for Advanced Studies of the Universe \& Department of Physics,\\
University of Illinois Urbana-Champaign, Urbana, IL 61801, USA}

\author{Justin Pickett}
\email{pickett.158@buckeyemail.osu.edu}
\affiliation{Department of Physics,
	         The Ohio State University, Columbus, OH\ 43210, USA}
	
\date{\today}

\begin{abstract}
A loosely bound hadronic molecule produced by a relativistic heavy-ion collision 
has been described as a ``snowball in hell'' since it emerges from a hadron resonance gas 
whose temperature is orders of magnitude larger than  the binding energy of the molecule.
This remarkable phenomenon can be explained in terms of a novel thermodynamic variable called the ``contact'' 
that is conjugate to the binding momentum of the molecule.
The production rate of the molecule  can be expressed in terms of the contact density at the kinetic freezeout of the hadron resonance gas. 
It approaches a nonzero limit as the binding energy goes to 0.
\end{abstract}


\maketitle
	
\textbf{Introduction.}
A molecule  in a medium whose temperature is much larger than the molecule's binding energy
is expected to disassociate almost immediately due to scattering with other particles.
However, loosely bound hadronic molecules \textit{have} been observed in heavy-ion collisions
and they seem to emerge from a hadron resonance gas
whose temperature is  orders of magnitude larger than the binding energy.
This problem has been referred to as the ``snowball in hell'' puzzle \cite{snowball}.
The phrase comes from the English idiom ``snowball's chance  in hell'', which refers to an event that is extremely unlikely.

According to the standard model of relativistic heavy-ion collisions \cite{Nagle:2018nvi,Elfner:2022iae},
a sufficiently central collision produces a region of {\it quark-gluon plasma}
that expands and cools and then transitions to a {\it hadron resonance gas}.
The hadron resonance gas expands and cools until {\it kinetic freezeout}, when it has become so dilute 
that  momentum distributions no longer change.
After  kinetic freezeout, the hadrons free stream to the detectors.
One might expect a hadronic molecule whose binding energy is much smaller than the energy scales at kinetic freezeout to have a snowball's chance in hell of surviving in the expanding hadron resonance gas.

 \begin{figure}[t]
 \includegraphics[width=0.48\textwidth]{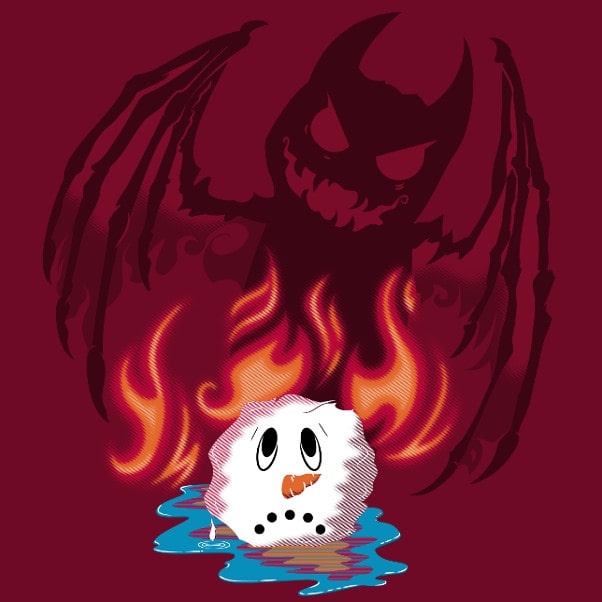}
  \caption{
A snowball in hell (image by Haneryuu).
 }
 \label{fig:snowball}
 \end{figure}

Several types of loosely bound hadronic molecules have been observed in heavy-ion collisions.
The STAR collaboration has observed the antideuteron \cite{STAR:2001pbk} 
and  the hypertriton \cite{STAR:2010gyg} in Au-Au collisions at the Relativistic Heavy Ion Collider.
The ALICE collaboration has observed the antideuteron  \cite{ALICE:2020zhb} 
and  the hypertriton \cite{ALICE:2022sco} in Pb-Pb collisions at the Large Hadron Collider (LHC).
The CMS collaboration has reported evidence for 
the charm-meson molecule $\chi_{c1}(3872)$ in Pb-Pb collisions at the LHC \cite{CMS:2021znk}.

Previous theoretical efforts to understand the production of loosely bound nuclei
in heavy-ion collisions have been reviewed in Refs.~\cite{Braun-Munzinger:2018hat,Oliinychenko:2020ply}.
A non-exhaustive list of recent efforts can be found in Refs.~\cite{Sombun:2018yqh,Kireyeu:2022qmv,Staudenmaier:2021lrg,Donigus:2022haq,Bratkovskaya:2022vqi,Wang:2022hja,Coci:2023daq,Cheng:2023rer,Knoll:2024gaf,Ege:2024vls}.
In simple {\it thermal models}, hadrons are produced in
 thermal equilibrium at the transition from the quark-gluon plasma to the hadron resonance gas \cite{Braun-Munzinger:2018hat}.
Thermal models have been remarkably successful in describing the rapidity distributions of loosely bound nuclei
as well as ordinary hadrons \cite{Cai:2019jtk}, but they cannot be easily applied to transverse momentum distributions.
In simple {\it coalescence models}, the momentum distribution of a molecule is proportional to
the product of the momentum distributions of its constituents \cite{BP-1963}.
The transverse momentum distributions of a molecule can be described by a sufficiently complicated coalescence model that depends on the detailed properties of the molecule,
e.g., Refs.~\cite{Nagle:1996vp,Scheibl:1998tk}.
Other models rely on dynamically generating the loosely bound states, such as the \textit{minimal spanning tree}~\cite{Bratkovskaya:2022vqi} or \textit{stochastic reactions}~\cite{Ege:2024vls}.

The \textit{contact}  is a thermodynamic variable relevant to systems whose constituents can form loosely bound molecules.
It was first introduced in cold-atom physics less than 20 years ago.
Here, we point out that the production rate of a loosely bound hadronic molecule in a heavy-ion collision 
is determined by the \textit{contact density} at kinetic freezeout.
We show how this observation can be exploited to predict the multiplicity of the molecule.

\textbf{Contact.}
The {\it contact} was introduced by Shina Tan in 2005 in the context of the {\it strongly interacting Fermi gas} \cite{Tan:0505,Tan:0508}, 
which consists of fermions with two spin states that interact only through an S-wave scattering length.
Tan derived a number of universal relations involving the contact that apply to any state of the system:
few-body or many-body, homogeneous or trapped, ground state or thermal, equilibrium or time-dependent.
The contact plays a central role in many experimental probes of ultracold atoms  \cite{Braaten:2010if}.
The first experimental verifications of Tan's universal relations using ultracold atoms were carried out in  2010 \cite{Jin-2010}.

The contact can be defined for any system that includes particles with an S-wave  scattering length $a$ 
that is large compared to the range of their interactions.
The few-body physics of these particles has universal aspects that are completely determined by $a$ \cite{Braaten:2004rn}.
We consider a system that includes two types of particles labeled by $\sigma=1,2$
with masses $m_1$ and $m_2$, reduced mass $m_{12}$, and  large scattering length $a$. 
If $a>0$, the two particles form a loosely bound molecule $X$ with binding energy $|E_X| = 1/(2 m_{12} a^2)$.
The molecule has a universal wavefunction $\exp(-r/a)/r$ for $r$ much larger than the range. 
Its constituents therefore have a large mean separation $r_X=a/2$.

In a many-body system, one might expect momentum distributions to fall off exponentially at large momentum $q$,
like the Fermi-Dirac or Bose-Einstein distributions.
Tan's {\it large-momentum relation} states that the momentum distributions $f_\sigma(\bm{q})$ 
of  the particles with a large scattering length have power-law tails \cite{Tan:0505}:
\beq
f_\sigma(\bm{q}) \longrightarrow C/q^4, \qquad \sigma = 1,2.
\label{n-k}
\eeq
The coefficient $C$, which is the same for both particles, is the {\it contact}.
We have normalized $f_\sigma(\bm{q})$ so the total number  of particles
of type $\sigma$ is $\int(d^3q/(2\pi)^3)f_\sigma(\bm{q})$. 
Note that the contact $C$, which has the dimensions of a momentum,
can take into account short-distance aspects of the interactions between the particles.

Tan's {\it adiabatic relation} expresses  the contact in terms of a derivative of the total energy $E$ of the system
with respect to the binding momentum $\gamma = 1/a$ at fixed entropy  \cite{Tan:0508}:
\beq
 C = -8 \pi m_{12} \left( \frac{\partial E}{\partial \gamma} \right)_{\!\!S}.
\label{C-dE/dgamma}
\eeq
Thus $-(C/8 \pi m_{12})d\gamma$ is the work done on the system 
by a small change in the scattering length.
The adiabatic relation implies that the contact is, up to a normalization factor, 
the extensive thermodynamic variable conjugate to $\gamma$.
In ultracold atoms, $\gamma$ can be controlled experimentally 
by tuning the magnetic field to near a Feshbach resonance \cite{Chin_2010}.
The contact density for a locally homogeneous system can be expressed as a derivative of the pressure at fixed temperature:
\beq
 \mathcal{C} = 8 \pi m_{12} \left( \frac{\partial P}{\partial \gamma} \right)_T.
\label{C-dP/dgamma}
\eeq
If we insert the energy  $E_X = -\gamma^2/2m_{12}$ for a single molecule into Eq.~\eqref{C-dE/dgamma},
we find that the contact for a single molecule $X$ is  $C_X = 8 \pi \gamma$.
In the dilute limit, the contact density is the sum of the contact for each particle and bound cluster weighted by their number densities. 
Since the only bound cluster in the strongly interacting Fermi gas is the loosely bound molecule, the contact density reduces to a single term given by the product of $C_X$ and the molecule number density $\mathfrak{n}_X$:
\beq
\mathcal{C}= 8 \pi \gamma\,  \mathfrak{n}_X.
\label{C-nX}
\eeq

\textbf{Virial expansion.}
If a system is sufficiently dilute, its thermodynamic variables can be calculated using the {\it virial expansion}.
We consider a homogeneous system in which the two particles with large scattering length $a$
are in thermal equilibrium at temperature $T=1/\beta$
and in chemical equilibrium with number densities  $\mathfrak{n}_\sigma$ are
determined by chemical potentials  $\mu_\sigma$.
If the number densities $\mathfrak{n}_\sigma$ are sufficiently low,
 the thermodynamic variables have virial expansions in powers of the fugacities  $z_\sigma= \exp(\beta \mu_\sigma)= \mathfrak{n}_\sigma (2\pi/m_\sigma T)^{3/2}$.

The leading term in the virial expansion for  the pressure from the interactions 
between the  particles of types 1 and 2 has the form
\bqa
 P_{12} = 2\,  T\, (m_{12} T/\pi)^{3/2}\,  b_{12} z_1 z_2 .
\label{P-virial}
\eqa
The interaction virial coefficient $b_{12}$ is a function of  $\gamma/ \sqrt{m_{12} T}$ only.
It can be deduced from the virial coefficient for identical particles
with 2-body phase shifts calculated by Beth and Uhlenbeck in 1937 \cite{BU-1937}:
\bqa
 b_{12} &=& \sqrt{2} \, e^{\beta \gamma^2/2m_{12}}\, \theta(\gamma)  
 \nonumber\\
&& -\frac{\sqrt{2}}{\pi} \int_0^\infty dp\,  \frac{\gamma}{\gamma^2 + p^2} \, e^{- \beta p^2/2m_{12}} .
\label{b2}
\eqa
The contact density can be obtained by differentiating the pressure as in Eq.~\eqref{C-dP/dgamma}
and then canceling the $\delta(\gamma)$ term by a subtraction term in the integral.
The leading term in the virial expansion is \cite{SL-2015} 
\beq
\mathcal{C} =  \frac{16}{\pi}\, (m_{12} T)^2\, z_1 z_2\,  F\big( \gamma/\sqrt{2m_{12}T}\,\big),
\label{Cdensity-F}
\eeq
where the dimensionless function $F(w)$ is 
\beq
F(w) = \frac{2w}{\sqrt{\pi}} \left( \pi\, e^{w^2}\, \theta(w)  
+  \int_0^\infty \!\!\!dx\,  \frac{x^2}{1+x^2}  \, e^{- x^2w^2} \right).
\label{F-z}
\eeq
It has an expansion in powers of $w$:  $F(w) = 1 + \sqrt{\pi} \, w + \ldots$ if $w>0$.

\textbf{Expanding hadron resonance gas.}
To derive a relation between the molecule number density and the contact density,
we use a toy model for the hadronic system produced by the  heavy-ion collision.
At a proper time $\tau$ after the collision, 
our toy model is a locally homogeneous system at a temperature $T(\tau)$ 
and a volume $V(\tau)$ \cite{Bjorken:1982qr}. 
At the transition from quark-gluon plasma to the hadron resonance gas, the system is in thermal and chemical equilibrium at a temperature $T_\mathrm{ch}\approx 156$~MeV \cite{Andronic:2017pug}.
After the transition, the system can be described by a decreasing temperature $T(\tau)$, an increasing volume $V(\tau)$, and a chemical potential $\mu_h(\tau)$ for each hadron $h$.  Each of the pions $\pi^+$, $\pi^0$, and $\pi^-$  
has decreasing number density $\mathfrak{n}_\pi(\tau)$.
The kinetic-freezeout temperature $T_\mathrm{kf}$ depends on the center-of-mass energy of the colliding ions and on the centrality of the collision.
After kinetic freeze-out, the short-lived resonances decay and resonances are no longer created by collisions. The hadron resonance gas near and after kinetic freezeout can alternatively be described by a hadron gas  consisting only of stable or nearly stable hadrons. The hadron gas can be obtained from the hadron resonance gas by integrating out the short-lived hadron resonances in favor of their decay products.

At the proper time $\tau_\mathrm{kf}$ of kinetic freezeout, the temperature of the hadron gas is $T_\mathrm{kf}$
and the total pion number density is $3\, \mathfrak{n}_{\pi \mathrm{kf}}$.
After kinetic freeze-out, the volume $V(\tau)$ 
continues to increase. The shapes of the momentum distributions of the hadrons remain the same as at kinetic freeze-out, where they are determined by $T_\mathrm{kf}$ and the hadron chemical potentials $\mu_{h \mathrm{kf}}$.
The number density $\mathfrak{n}_h(\tau)$ of  hadron $h$ decreases in proportion to  $1/V(\tau)$.
The ratio $\mathfrak{n}_h(\tau)/\mathfrak{n}_{\pi}(\tau)$ of the number densities of the hadron and a pion therefore
remains fixed and  must be equal to the ratio of the multiplicities $dN/dy$ observed at the detector.

\textbf{Evolution of the contact density.}
Ordinary hadrons have strong nuclear interactions 
whose range is comparable to or shorter than the inverse pion mass $1/m_\pi = 1.41$~fm.
They are therefore essentially noninteracting after kinetic freezeout.
The constituents of a loosely bound hadronic molecule $X$ are exceptions. 
After kinetic freezeout,  they decouple from the pions and other hadrons
but they continue to interact with each other through their small binding momentum $\gamma_X$.
They can be described by an effective field theory near the unitary renormalization-group fixed point 
defined by $\gamma_X=0$.
This RG fixed point is a nonrelativistic conformal field theory \cite{Nishida:2007pj}.
 The contact density $\mathcal{C}_X(\tau)$ associated with $X$ is the expectation value of an operator with scaling dimension 4 \cite{Braaten:2008uh}, so it decreases as $V(\tau)^{-4/3}$ or equivalently $\mathfrak{n}_\pi(\tau)^{4/3}$.
This anomalous scaling behavior with exponent 4/3 continues until a time $\tau_X$ when there is a crossover to the conventional scaling behavior with exponent 1:
\beq
\mathcal{C}_X(\tau) = \mathcal{C}_X(\tau_\mathrm{kf} ) 
\left[ \mathfrak{n}_{\pi}(\tau) / \mathfrak{n}_\pi(\tau_\mathrm{kf}) \right]^{4/3},  ~ \tau_\mathrm{kf} < \tau \lesssim \tau_X.
\label{C-tau}
\eeq
In the dilute limit, the contact density is given by Eq.~\eqref{C-nX}:
$\mathcal{C}_X(\tau) = 8 \pi \gamma_X\,  \mathfrak{n}_X(\tau)$,
where $\mathfrak{n}_X(\tau)$ is the molecule number density. The conventional scaling behavior required by the dilute limit is proportional to $1/V(\tau)$ or equivalently $\mathfrak{n}_\pi(\tau)$:
\beq
\mathcal{C}_X(\tau) = \mathcal{C}_X(\tau_X ) 
\left[ \mathfrak{n}_{\pi}(\tau)/  \mathfrak{n}_\pi (\tau_X)  \right],   \quad \tau \gtrsim \tau_X.
\label{C-tau:dilute}
\eeq
Eq.~\eqref{C-tau:dilute} is the familiar statement that as the volume 
$V(\tau)$ of a system of non-interacting particles  increases, their number densities scale as $1/V(\tau)$.
Eq.~\eqref{C-tau} is the analogous statement for interacting particles in a quantum field theory near a nontrivial renormalization-group fixed point.
Densities scale in proportion to $V(\tau)$ raised to the appropriate anomalous dimension.
After inserting Eq.~\eqref{C-tau:dilute} for $\mathcal{C}_X(\tau)$ and then using Eq.~\eqref{C-tau} for
$\mathcal{C}_X(\tau_X)$,  the molecule number density for $\tau > \tau_X$ reduces to
\beq
\mathfrak{n}_X(\tau) = \frac{1}{8 \pi \gamma_X}\,  \mathcal{C}_X(\tau_\mathrm{kf}) 
\left( \frac{\mathfrak{n}_\pi(\tau_X)}{\mathfrak{n}_\pi(\tau_\mathrm{kf})} \right)^{\! 1/3} \frac{\mathfrak{n}_\pi(\tau)}{\mathfrak{n}_\pi(\tau_\mathrm{kf})} .
\label{nX-C}
\eeq
As an estimate of the crossover time $\tau_X$ in the hadron resonance gas, 
we take the time when the mean distance  $r_\pi(\tau)$ to the nearest pion
exceeds the mean separation $r_X = 1/(2 \gamma_X)$ of the constituents of the molecule
by a numerical factor:
$r_\pi(\tau_X) = \big[ \Gamma(\tfrac43)\,(4\pi)^{-1/3}/\kappa \big]r_X$.
We expect the coefficient of $r_X$
to be roughly 1, but we will treat $\kappa$ as a phenomenological parameter.
The mean pion distance in a homogeneous system with uniform pion number density $\mathfrak{n}_{\pi}(\tau)$ is $r_\pi(\tau)=\Gamma(\tfrac43)[4 \pi \mathfrak{n}_{\pi}(\tau)]^{-1/3}$ \cite{Hertz1909}. 
The pion number density at $\tau_X$ then  reduces to
$\mathfrak{n}_{\pi}(\tau_X) = (2\kappa \gamma_X)^3$.
This simple expression explains why we chose the complicated
expression for $r_\pi(\tau_X)$ above. Given that $\mathfrak{n}_{\pi}(\tau)$ after kinetic freezeout scales roughly as $1/\tau^3$, the crossover time $\tau_X$ scales as $1/\gamma_X$.
Since the ratio of the number densities of $X$ and $\pi$ for $\tau > \tau_X$ is equal to the ratio of their multiplicities,
the multiplicity of the molecule  is
\beq
dN_X/dy = \frac{ \kappa}{4 \pi} \left( \mathcal{C}_{X\mathrm{kf}} / \mathfrak{n}_{\pi \mathrm{kf}}^{4/3} \right)  dN_\pi/dy ,
\label{dN/dy-X}
\eeq
where $\mathcal{C}_{X\mathrm{kf}}$ and $\mathfrak{n}_{\pi \mathrm{kf}}$ 
are the contact density and pion number density at kinetic freezeout.
This expression for the multiplicity of the molecule is the primary result of our paper. 
It does not depend on the toy model used in its derivation.
Note that $dN_X/dy$ depends on $\gamma_X$ only through the contact density at kinetic freezeout. 
Since $\mathcal{C}_\mathrm{Xkf}$ has a nonzero limit as $\gamma_X \to 0$, the multiplicity  in Eq.~\eqref{dN/dy-X} is nonzero in that limit.
This disagrees with the intuition that the production rate of a loosely bound molecule should go to 0 as its binding energy goes to zero. 

To illustrate the application of Eq.~\eqref{dN/dy-X} for the molecule multiplicity, we approximate the contact density $\mathcal{C}_{X\mathrm{kf}}$ at kinetic freezeout by the leading term in the virial expansion 
for its constituents in Eq.~\eqref{Cdensity-F} evaluated at $T = T_\mathrm{kf}$. The number densities $\mathfrak{n}_1$ and $\mathfrak{n}_2$ of the constituents are those in a hadron gas in which short-lived resonances have been integrated out.  The multiplicity of a molecule $X$ with binding momentum $\gamma_X$ is then
\bqa
dN_X/dy &=&  32 \pi\,  \kappa\, f_X\,
F\big( \gamma_X/  \sqrt{2 m_{12}T_\mathrm{kf}} \big) 
\nonumber\\
&&  \hspace{-0.5cm}\times 
\left( \!\frac{m_{12}\, \mathfrak{n}_{\pi \mathrm{kf}}^{4/3}}{m_X^3 T_\mathrm{kf}^2 }\! \right)^{\! 1/2} 
\frac{dN_1/dy\, dN_2/dy}{dN_\pi/dy}.~~~
\label{multiplicity-X}
\eqa
Since $m_h \gg T_\mathrm{kf}$, we have used the Boltzmann approximation to express the fugacity for a constituent hadron $h$ as
$z_h= \mathfrak{n}_h(\tau_\mathrm{kf})\,  (2\pi/m_hT_\mathrm{kf})^{3/2}/(2s_h \!+\!1)$, where $s_h$ is the hadron spin.
The factor $f_X$ in Eq.~\eqref{multiplicity-X} is the fraction of the scattering channels for the two constituents that have the small binding momentum $\gamma_X$.
We have replaced the ratios $\mathfrak{n}_h(\tau)/\mathfrak{n}_{\pi}(\tau)$ by the corresponding ratios of  multiplicities.
We have also replaced $m_1 \!+\! m_2$ by the mass $m_X$ of the molecule.

\textbf{Estimation of $\bm{\kappa}$.}
The deuteron ($d$) is a proton-neutron $(pn)$ bound state with spin 1, isospin 0, and a relatively small binding energy 2.225~MeV.
The fraction of $np$ scattering channels that are resonant is $f_d = 3/8$.
The ALICE collaboration has measured the production of the deuteron (and antideuteron) in Pb-Pb collisions
at the center-of-mass energy per nucleon  $\sqrt{s_{NN}}=2.76$~TeV \cite{ALICE:2015wav}.
The mean deuteron multiplicity $dN_d/dy$  in the 0-10\% bin of the centrality of the collision
is $(9.8 \pm 1.6) \!\times\! 10^{-2}$.
We will use this result to obtain an order-of-magnitude estimate of the parameter $\kappa$ in Eq.~\eqref{dN/dy-X}. 

The mean proton multiplicity $dN_p/dy$ (which is equal to that of the neutron by isospin symmetry)
and the mean pion ($\pi^+$ or $\pi^-$) multiplicity $dN_\pi/dy$ were measured in the 0-5\% and 5-10\% centrality bins in Ref.~\cite{ALICE:2013mez}. 
The temperature $T_\mathrm{kf}$ at kinetic freezeout
can be estimated through blast-wave fits to the transverse momentum distributions of $\pi$, $K$, and $p$.
For Pb-Pb collisions at $\sqrt{s_{NN}} = 2.76$\,TeV, a fit that also allows for a pion chemical potential $\mu_{\pi\mathrm{kf}}$ has given $T_\mathrm{kf} = 78.3 \pm 1.6$ MeV and $\mu_{\pi\mathrm{kf}} \approx 90$~MeV \cite{Melo:2019mpn}.
Upon inserting  these results into Eq.~\eqref{multiplicity-X} and solving for $\kappa$, we obtain 
$\kappa_d = 0.18 \pm 0.04$,
where the errors from  multiplicities have been combined in quadrature.

It is useful to have quantitative estimates of the relevant length scales.
The mean pion distance at kinetic freezeout with $T_\mathrm{kf} = 78.3$\,MeV and $\mu_{\pi\mathrm{kf}} = 90$~MeV is $r_{\pi\mathrm{kf}}  =1.61$~fm.
This is  a little smaller than the mean separation of the constituents of the deuteron inferred from its binding energy:
$r_d = 2.16$~fm. 
Our estimate for $\kappa_d$ implies that the mean pion distance at the deuteron crossover time is $r_\pi(\tau_d) \approx 2.1\, r_d$.

\textbf{Multiplicities.}
Eq.~\eqref{dN/dy-X} implies that the ratio of the multiplicities of two loosely bound molecules is just the ratio of the corresponding contact densities at kinetic freezeout.  If the contact density is approximated by the leading term in the virial expansion in Eq.~\eqref{Cdensity-F}, the ratio for a loosely bound molecule $X$ and the deuteron is
\bqa
&&\frac{dN_X/dy}{dN_d/dy} =  \frac{f_X}{f_d}
\left( \frac{m_{d}^3\, m_{12}}{m_{X}^3\, m_{pn}} \right)^{\!1/2}
\nonumber\\
&&  \hspace{.2cm}\times
\frac{F\big( \gamma_X/  \sqrt{2 m_{12}T_\mathrm{kf}} \big) }{F\big( \gamma_d/  \sqrt{2 m_{pn}T_\mathrm{kf}} \big) }
\frac{(dN_1/dy)(dN_2/dy)}{ (dN_p/dy)^2}.
\label{multiplicity-ratio}
\eqa
The contact density at kinetic freezeout 
enters only through the ratio of the functions $F(w)$.
The function in the numerator approaches 1 as $\gamma_X \to 0$.

The hypertriton ($^{3\!}_{\Lambda\!}$H) is a $p n \Lambda$ bound state with  spin $\frac12$ that is essentially 
a molecule composed of a  deuteron and the strange baryon $\Lambda$. 
The fraction of $d\, \Lambda$ scattering channels that are resonant is $f_{^{3\!}_{\Lambda\!}\mathrm{H}} = 1/3$.
The $\Lambda$ separation energy has been measured in emulsion experiments \cite{Juric:1973zq}
and in heavy-ion collisions \cite{STAR:2019wjm,ALICE:2022sco}. The average of the existing measurements of the $\Lambda$ separation energy is $148\pm40$~keV 
\cite{Eckert:2022srr}.
The ALICE collaboration has measured the production rate of the hypertriton (and anti-hypertriton)
in Pb-Pb collisions at $\sqrt{s_{NN}}=2.76$~TeV \cite{ALICE:2015oer}.
The mean multiplicity in the 0-10\% centrality bin multiplied by the branching fraction
of the hypertriton into $^{3\!}\mathrm{He}\, \pi^-$  is $(3.67 \pm 0.74) \!\times\! 10^{-5}$,
where the errors have been added in quadrature. 
The  mean hypertriton multiplicity in that bin can be predicted  by inserting
the multiplicities for $p$, $d$, and $\Lambda$ from Refs.~\cite{ALICE:2013mez}, \cite{ALICE:2015wav}, and \cite{ALICE:2013cdo} 
into Eq.~\eqref{multiplicity-ratio} along with $T_\mathrm{kf}= 78.3$~MeV.
In our approximation for the contact density, we ignore the fact that the deuteron number density at kinetic freezeout may not be well-defined. 
It may be possible instead to express the contact density associated with the hypertriton in terms of the contact density associated with the deuteron, which is well defined at kinetic freezeout.
Our prediction for the hypertriton multiplicity is $(10.4 \pm 3.9) \!\times\! 10^{-5}$, 
where the errors from multiplicities have been combined in quadrature.
The prediction is insensitive to the $\Lambda$ separation energy.
Our prediction multiplied by the 25\% branching fraction into $^{3\!}\mathrm{He}\, \pi^-$ \cite{Kamada:1997rv} is  consistent with the ALICE result to within the errors.

The $\chi_{c1}(3872)$ is  a loosely bound charm-meson molecule discovered in 2003 \cite{Belle:2003nnu}.
The difference between its mass and the threshold for the charm-meson pair $D^{*0}\bar D^0$
is  $-50 \pm 93$~keV \cite{PDG2004}. 
Its quantum numbers $J^{PC} = 1^{++}$ \cite{LHCb:2013kgk} imply that  its constituents 
are the linear combination $D^{*0} \bar D^0 + D^0 \bar D^{*0}$.
The fraction of $D^{*0} \bar D^0$ and $D^0 \bar D^{*0}$ scattering channels that are resonant is $f_X = 1/2$.
The CMS collaboration has presented evidence for the production of $\chi_{c1}(3872)$
in Pb-Pb collisions  at $\sqrt{s_{NN}}=5.02$~TeV  \cite{CMS:2021znk}.
In order to use Eq.~\eqref{multiplicity-ratio} to predict the multiplicity of $\chi_{c1}(3872)$, 
we need the multiplicities of its charm-meson constituents, which have not been measured.
They have however been predicted for Pb-Pb collisions  at $\sqrt{s_{NN}}=5.02$~TeV 
using the Statistical Hadronization Model with charm quarks (SHMc) \cite{Andronic:2021erx}.
The mean $D^{*0}$  multiplicity  should by isospin symmetry be equal to that for $D^{\ast +}$:
$2.4 \pm 0.4$  in the 0-10\% centrality bin.
The mean multiplicity for $D^0$ before $D^*$ decays can be inferred by isospin symmetry
from the SHMc predictions for $D^+$ and $D^{*+}$: $1.9 \pm 0.5$  in the 0-10\% bin. 
The mean proton multiplicity in the 0-10\% bin is given in Ref.~\cite{ALICE:2019hno}.
The mean deuteron multiplicity in the 0-10\% bin
can be obtained from Fig.~4 of Ref.~\cite{ALICE:2022veq}:
$dN_d/dy=(11.9\pm 0.4)\times 10^{-2}$.
Inserting these results into Eq.~\eqref{multiplicity-ratio} along with $T_\mathrm{kf}= 78.3$~MeV,
our prediction for the mean multiplicity of $\chi_{c1}(3872)$ in the 0-10\% bin is  $(23.4 \pm 7.8) \!\times\!10^{-5}$,
where the errors  from multiplicities have been combined  in quadrature.
The predicted multiplicity is insensitive to the $\chi_{c1}(3872)$ binding energy.

\textbf{Conclusions.}
We have shown that the production rate of a loosely bound hadronic molecule in relativistic heavy-ion collisions
can be determined from the contact density of the resulting hadron  gas at kinetic freezeout using Eq.~\eqref{dN/dy-X}.
We illustrated the application of  that equation by approximating the contact density 
by the leading term in the virial expansion for the constituents.
The resulting expression for the ratio of the multiplicities of a loosely bound hadronic molecule and the deuteron is given in Eq.~\eqref{multiplicity-ratio}.
We then used this equation to calculate
the multiplicities of the hypertriton and  the $\chi_{c1}(3872)$.
The predicted hypertriton multiplicity is consistent with the measured value to within the errors. 

 The contact density can be calculated using any model for the hadron gas in which the loosely bound molecule 
is generated dynamically by the large scattering length of its constituents. 
An important step would be to improve on our simple approximation using the virial expansion to obtain a more accurate description of the contact density in the hadron gas.
Our approximation for the contact density does not take into account 3-body effects.
The most important  3-body effects involve pions, since they are the most abundant constituents of the hadron gas.
Our approximation for the contact density
could be improved by using an effective field theory near the unitary fixed point that includes pions 
as well as the constituents of the molecule \cite{Kaplan:1998we,Fleming:2007rp}.
The contact density can be calculated as an expansion in the fugacities of the heavy constituents and in the coupling constant for their interactions with  pions.

We have related the abundance of a loosely bound 2-body hadronic molecule in heavy-ion collisions to the contact density at kinetic freezeout. In addition to the loosely bound 2-body molecule, there may be loosely bound 3-body molecules, in which case the system may have universal properties determined by the 3-body contact \cite{Braaten:2011sz}. 
It would be interesting to see if these universal properties can be applied to the abundance of loosely bound 3-body hadronic molecules, such as H$^3$ and He$^3$ nuclei, in heavy-ion collisions.

Given our approximation for the contact density, we calculated the multiplicities of  loosely bound molecules.
Our methods can be extended to calculate the transverse momentum distribution. 
The blast-wave fits of hadron momentum distributions  determine not only $T_\mathrm{kf}$
but also the phase-space distribution of the hadronic fluid.
The transverse momentum distribution of a molecule can be predicted by assuming it has a Maxwell-Boltzmann distribution in the co-moving frame of the expanding fluid at the crossover time.

Our methods can be applied to other loosely bound hadronic molecules,
such as $T_{c c}^+(3875)$,  a charm-meson molecule with constituents $D^{*+} D^0$
and binding energy  $273 \pm 62$~keV  discovered in 2021 \cite{LHCb:2021vvq}.
They can also be applied to the production of loosely bound molecules in other fields of physics.
Cold-atom physics provides  systems that are theoretically pristine. 
The scattering length can be controlled experimentally
and used to make the binding energy of a molecule arbitrarily small.
Systems  in which an atomic gas escapes from a trapping potential  can be engineered with exquisite experimental control. 
They should allow quantitative studies of the role played by the contact in the production of snowballs from hell.

\begin{acknowledgments}
This work was supported in part by the U.S.\  Department of Energy under grant DE-SC0011726
and by the National Science Foundation under Grant NSF PHYS-2316630.
This work contributes to the goals of the US DOE ExoHad Topical Collaboration, Contract DE-SC0023598.
KI would like to thank J.~Noronha and J.~Noronha-Hostler for their feedback on and discussion of the manuscript.
\end{acknowledgments}

\end{document}